\documentclass[nofootinbib,twocolumn,preprintnumbers,superscriptaddress]{revtex4-1}
\usepackage{amsmath}
\usepackage{amssymb}
\usepackage{graphicx}
\usepackage{hyperref}
\usepackage{xspace}
\usepackage{bm}
\usepackage{slashed}
\usepackage{dsfont}
\usepackage{bbm}
\usepackage{esint}
\usepackage{epstopdf}

\usepackage{booktabs}
\AtBeginDocument{
\heavyrulewidth=.08em
\lightrulewidth=.05em
\cmidrulewidth=.03em
\belowrulesep=.65ex
\belowbottomsep=0pt
\aboverulesep=.4ex
\abovetopsep=0pt
\cmidrulesep=\doublerulesep
\cmidrulekern=.5em
\defaultaddspace=.5em
}

\newcommand{\be}{\begin{equation}}
\newcommand{\ee}{\end{equation}}
\newcommand{\bea}{\begin{eqnarray}}
\newcommand{\eea}{\end{eqnarray}}
\newcommand{\en}{\mathcal{E}}
\usepackage{xcolor}
\definecolor{light-gray}{gray}{0.8}

\begin{document}

\title{Axion-like particle searches with sub-THz photons}


\newcommand{\CERNaff}{CERN-TH, CH-1211 Geneva 23, Switzerland}
\newcommand{\sapienza}{Dipartimento di Fisica and INFN, `Sapienza' Universit\`a di Roma\\
P.le Aldo Moro 5, I-00185 Roma, Italy}
\newcommand{\nest}{ NEST, Istituto Nanoscienze-CNR and Scuola Normale Superiore, I-56127 Pisa, Italy }
\newcommand{\infnroma}{INFN Sezione di Roma, P.le Aldo Moro 5, I-00185 Roma,  Italy}
\newcommand{\infnpisa}{INFN Sezione di Pisa,  Largo Bruno Pontecorvo, 3, 56127 Pisa, Italy}
\newcommand{\UCLA}{Department of Physics and Astronomy, University of California Los Angeles,\\ 475 Portola Plaza, Los Angeles, CA 90095, USA}

\author{L.M.~Capparelli}
\affiliation{\UCLA}

\author{G.~Cavoto}
\affiliation{\infnroma}

\author{J.~Ferretti}
\affiliation{\sapienza}

\author{F.~Giazotto}
\affiliation{\nest}

\author{A.D.~Polosa}
\affiliation{\sapienza}
\affiliation{\CERNaff}

\author{P.~Spagnolo}
\affiliation{\infnpisa}

\begin{abstract}
We propose a variation, based on very low energy and extremely intense photon sources, on the well established technique of Light-Shining-through-Wall (LSW) experiments for axion-like particle searches.  With radiation sources at 30 GHz, we compute that present laboratory exclusion limits on axion-like particles might be improved by at least four orders of magnitude, for masses $m_a\lesssim 0.01$~meV.  This could motivate research and development programs on dedicated single-photon sub-THz detectors.   

 \end{abstract}

\pacs{-----}

\maketitle
\section{Introduction}
There is a vast literature describing the theoretical motivations for axion-like particles and the experimental techniques to investigate on their existence~\cite{gen}. The topic has received a renovated attention because axions may well be among  dark-matter constituents.

The purpose of this letter is to discuss a variation on the well-established technique of Light-Shining-through-Wall (LSW) experiments. We suggest to use
\begin{enumerate}
\item Very intense sub-THz sources: gyrotrons at frequencies $\nu\approx 30$~GHz.
\item Single photon detectors for light in  this frequency domain.
\end{enumerate}

Gyrotron sources can provide photons from few  to  several 100~GHz. $\nu\approx 30$~GHz is a value which allows to avoid the strong ambient microwave background still ensuring  an extremely  large photon yield\footnote{In tested gyrotrons, the average beam radius is $\sim 1.3$~cm and the average axial spread is about $5.5 \%$. The bandwidth is of the order of 30 MHz.}.

In a typical LSW experiment, the rate of events $\dot N_{\rm evts}$ is driven by $\dot N_\gamma$, the rate of photons delivered by the source 
\be
\dot N_{\rm evts}\propto \dot N_\gamma\, P_{\gamma\to a}\times P_{a\to\gamma}\sim\dot N_\gamma\, G^4 H^4 L^4
\ee
where $G$ is the (unknown) photon-axion coupling, $H$ is the strength of an external magnetic field and $L$ is the  length of the photons path  under the action of $H$. $P_{\gamma\to a}=P_{a\to \gamma}$ is the probability of photon-to-axion-conversion (and viceversa).

If $G\lesssim 10^{-10}$~GeV$^{-1}$, as suggested by CAST \cite{Arik:2015cjv}, and an external magnetic field  $H=15$~T is devised along $50$~cm, we have that  $(GHL)^4\lesssim 10^{-35}$. 

Mega-Watt gyrotron sources can produce over $\dot N_\gamma\approx 10^{28}$ photons/second with continuous emission
(in the same conditions a 1~W  LASER source yielding visible photons can reach $\dot N_\gamma\approx 10^{18}$ where 1~W~$=6\times 10^{18}$~eV/sec). This would allow $\approx 10$ LSW events per year at $G\sim 10^{-10}$~GeV$^{-1}$. 

The number of expected events will grow by a factor of $Q$ if a Fabry-Perot resonant cavity (for $\approx$~30~GHz photons) with quality factor $Q$ encloses the magnetic field region where photon-to-axion conversion is expected to occur.   Quality factors of Fabry-Perot cavities in the microwave domain are known to be way larger than those  typical for optical radiation ($Q=10^4\div 10^5$ vs 50, see for example~\cite{crq}). A   gyrotron with a bandwidth matching  the   cavity quality factor must be eventually  chosen  for  the real experimental setup.

On the other hand, detectors sensitive to a single photon  in the frequency range around $30$~GHz are needed, especially if we want to push the exclusion limits below $G\sim 10^{-10}$~GeV$^{-1}$.  Such devices 
need dedicated research and development along the lines we will describe below.

\begin{figure}[htb!]
\begin{center}
\includegraphics[width=1.0\columnwidth]{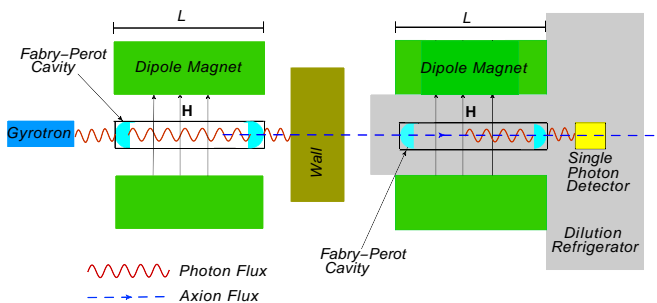}
\end{center}
\caption{\small Scheme of the LSW experiment with low energy photons. The dilution fridge should extend inside the dipolar magnet to easily accept photons from axion conversion. Such region would be thermally anchored to the cryostat mixing chamber residing at the base temperature of 10 mK.
The regeneration cavity - if any - would then be placed in the fridge itself as well.  }   
\label{zero}
\end{figure}

If detectors of this kind will be engineered, low-energy LSW experiments would be far more reaching at probing low values of $G$ couplings than those based on  LASER sources. 
We include in Fig.~\ref{zero} a qualitative scheme of  the experimental setup we propose.

On the theory side, we expect that if the energy of photons from the source is lowered down to the (unknown) mass value of the axion, we might be producing large numbers of axions at rest. We will discuss about how to compute photon-axion-photon conversion probabilities as $\en_\gamma$ approaches the $\en_\gamma = m_a$ limit.   

Indeed, photons at about 30~GHz have $\approx  10^{-4}$~eV, so that $\en_\gamma$ in principle might hit the  $m_a$ axion mass value; this is never the case for standard LSW configurations with visible light sources, where the approximation $\en_\gamma\gg m_a$ is perfectly appropriate given the existing experimental limits on $m_a$. 

Formulas describing photon-axion conversion rates are known to contain a factor of ${\mathcal F}=\en_\gamma/\sqrt{\en_\gamma^2-m_a^2}$, which is typically set to $1$ but might be of concern at  very low photon energies. 

In the next Section, we will discuss a formula for the conversion probability, which is also valid in the proximity of the $\en_\gamma=m_a$ limit. 
We will conclude that the extension $L$ of the magnetic region (in the photon beam direction) affects the denominator in the factor $\mathcal{F}$, shifting it to \mbox{$1/L+\sqrt{\en_\gamma^2-m_a^2}$}.

Next, in a dedicated section, we will discuss the potentialities of a LSW experiment with 30~GHz photons, showing how far it could improve the actual exclusion limits. Paraphotons and chameleons searches will be addressed separately. We conclude with a discussion on the feasibility of single-photon microwave detectors.

During the preparation of this paper we learned
that there are some ongoing  studies on LSW with gyrotrons, although in a rather different energy range~\cite{japgyr}. 

\section{Low Energy Photon-Axion Conversion}
By axion-like particles one generally means the quanta of a massive scalar or pseudoscalar field $a$ coupled to photon $\bm A$ by an interaction
\be
\mathcal{L}_I=\frac{1}{4}G\,a\,F^{\mu\nu}\tilde{F}_{\mu\nu}=G\, a\,\bm H\cdot \partial_0\bm A
\label{elle}
\ee
Here, without loss of generality, the pseudoscalar case is considered. The coupling $G$ has dimensions of the inverse of a mass and is a free parameter. The external field $\bm H$ is chosen to be uniform in space and constant in time (static). 

A real photon traveling in the $x$ direction within an external magnetic field orthogonal to it has a finite probability to convert into an axion-like particle (ALP or axion for brevity)  by exchanging 3-momentum $\bm q$ with $\bm H$, while conserving the energy: $\en_\gamma=\en_a$. 

Taking $\bm q$ along the photon beam direction (the $x$ direction) allows $p^2>0$, $p$ being the axion 4-momentum.

The inverse process is also possible: the axion has a finite probability to convert back into a photon carrying the same energy as the original one, and collinear to it.   

The external static magnetic field is assumed to be uniform and equal to $H$ within a volume $L_xL_yL_z=L_x S$, where $L_x$ is along the photon beam direction and $S$ is the effective transverse section of the region permeated by $\bm H$, crossed by the photon beam.  

One generally analyzes  the contribution to the photon-axion conversion S-matrix, where only $q_x$ is exchanged with the magnetic field. 
Integrating ${\cal L}_I$ in $d^4x$, we get the folowing $\langle \gamma\gamma|a\rangle$ matrix element\footnote{From~(\ref{elle}), the Feynman rule at the photon-axion vertex for an incoming/outgoing photon is 
$ iG\, \bm H\cdot\bm \epsilon(k)(\mp i\mathcal{E}_\gamma)$,
where $\bm \epsilon(k)$ is the transverse photon polarization vector. 
For a transverse photon, the sum over polarizations is  
\mbox{$\sum_\lambda |\bm H\cdot\bm \epsilon(k,\lambda)|^2= H^2\sin^2\theta=H^2$}, 
$\theta=\pi/2$ being the angle between the photon direction  and the orientation of the external magnetic field. }
\be
G H\, (2\pi)^3\delta(q_y)\delta(q_z)\delta(\en_a-\en_\gamma)\,2\frac{\sin(q_x L_x/2)}{q_x}\, (\pm \mathcal{E}_\gamma) 
\label{frule2}
\ee
depending if the photon is incoming/outgoing. 
The $2 \sin(q_xL_x/2)/q_x$ factor comes from taking the integration range in the $x$ variable in the finite interval $[-L_x/2,L_x/2]$, representing the extension of the external magnetic field along the longitudinal photon beam direction. 

In the limit $L_x\to \infty$, we have $2 \sin(q_xL_x/2)/q_x\to 2\pi\delta(q_x)$. However, $q_x=0$ would be equivalent to $p_x=k_x$. Therefore, the energy delta-function would be zero for $m_a\neq 0$, {\it i.e.} no photon-axion transition would be allowed. 

The conservation of energy derives from the fact that the external magnetic field is constant in time. 

Using standard normalizations for photon and scalar wave functions in a box~\cite{weinbergokun}, 
the transition probability $\gamma\to a$ is
\bea
dw_{a\gamma}&=&G^2H^2\frac{(2\pi)^3}{(2\en_aV) (2\en_\gamma V)}\delta(q_y)\delta(q_z) \delta(\en_a-\en_\gamma)\times \notag\\
&\times& (ST)\, 4\frac{\sin^2(q_x L_x/2)}{q_x^2}\, \mathcal{E}_\gamma^2\,\frac{V}{(2\pi)^3}d^3 p
\label{damod}
\eea
where $S\equiv (2\pi \delta(0))^2$ is the magnetic field transverse section introduced above and $2\pi \delta(0)=T$ for the energy  delta-function.  $V=L_x\cdot S$ is the volume with $H\neq 0$. 

Considering that $\en_a=\sqrt{p_x^2+m_a^2}$ and $q_y=p_y$, $q_z=p_z$, solving  for $p_x>0$ only, with  photon polarizations being fixed,  we get the conversion rate per  flux (as in~\cite{gen} and~\cite{vanbibbr})
\bea
\sigma=\frac{1}{\rho v}\int\frac{dw}{T}= S G^2H^2 \, \frac{\sin^2(q_x L_x/2)}{q_x^2}\,\frac{\en_\gamma}{\sqrt{\en_\gamma^2-m_a^2}}
\label{rate1}
\eea
where $q_x=\en_\gamma-\sqrt{\en_\gamma^2-m_a^2}$ and $\rho v={\rm flux}=1/V$.
This is the result generally used in the literature.

The  factor $\en_\gamma/\sqrt{\en_\gamma^2-m_a^2}$ is usually set to $1$ because it is assumed that $\en_\gamma\gg m_a$ (in this case we also have $q_x\approx m_a^2/2\en_\gamma \neq 0$) -- most of the present experiments work in that regime.

However, when decreasing $\en_\gamma$ in the sub-THz  regions we want to explore, it might be that $\en_\gamma\approx m_a$. This formula would indicate that in the $\en_\gamma\to m_a$ limit, the production probability dramatically increases for the production of axions at rest (which are of no use in a LSW experiment).

The photon energy regime we are mostly interested in is  $\mathcal{E}_\gamma\approx 10^{-4}$~eV. Therefore, we can safely compute reaction rates through formula~(\ref{rate1}) whenever axion masses are $\lesssim 0.1$~meV. 

\subsection*{An alternative derivation}
In what follows, we wish to discuss the $\en_\gamma=m_a$ limit  from a different standpoint.  We will obtain a simple way of regulating the expression in~(\ref{rate1}).

\begin{figure}[h]
\begin{center}
\includegraphics[width=1.0\columnwidth]{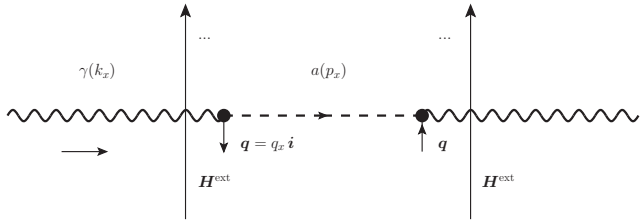}
\end{center}
\caption{\small Photon-axion-photon conversion in an external magnetic field orthogonal to the photon direction; $q_x$ can  either be positive or negative. This scheme does not describe the typical LSW photon conversion: here the photon-axion-photon conversion occurs within the same transverse magnetic field area, along a region of arbitrary length $L$.}   
\label{uno}
\end{figure}

The diagram represented  in Fig.~\ref{uno} might  be considered as a sort of photon  self-energy  contribution $\Sigma(k)$ once the action of the external magnetic field is substituted by the emission and re-absorption of  arbitrary values of  the 3-momentum $\bm q$ along $x$.  We stress here that the 
physical situation we are studying now does not correspond to the LSW photon-axion-photon conversion for there is no wall in the scheme in  Fig.~\ref{uno}. The scope of this Section is to explicitly show  how the scale $1/L$ plays the role of an infrared  cutoff in the formula commonly used for the photon-axion conversion probability -- something which we might have guessed on dimensional grounds only. 

Another diagram  will implicitly be included in which the photon converts into a bouncing axion against the magnetic wall, eventually regenerating a photon collinear to the original one (upon a second reflection in $H$). 

Indeed, if an interaction with axion-like particles of the kind of Eq.~(\ref{elle}) exists, photons in external magnetic fields become unstable with respect to photon-to-axion decay and the conversion probability is expected to be proportional to the imaginary part of the self-energy diagram(s) in Fig.~\ref{uno} (including the bouncing axion).  

We will compute $\Sigma(k)$ in Fig.~\ref{uno}, where the role of the external magnetic field is that of a sink/source of 3-momentum. Then, we use unitarity (cutting equations\footnote{We use $2\,\mathrm{Im}\,A(k\to k)=\sum_f\int \prod^f (d^3 p_f/(2\pi)^3 2E_f)  |A(k\to f)|^2 $.}) to re-obtain the rate of photon conversions in the external magnetic field
\be
P=\frac{\mathrm{Im}(\Sigma)}{\en_\gamma}
\label{forml}
\ee
using the notations of Eq.~(\ref{rate1}),  $P=P_{\gamma \to a}=P_{a \to \gamma}=\sigma/S$, and including the appropriate wave function normalization factors $1/(\sqrt{2\en_\gamma})^2$.

From now on, we will simplify the notation by setting $L_x=L$, $p_x=p$, $k_x=k$, $q_x=q$. 

As shown in the Appendix, we obtain for the self-energy 
\be
\Sigma(k)=-4\,G^2 H^2\,\mathcal{E}_\gamma^2\, \int\,\frac{dq}{2\pi} \frac{\sin^2(q L/2)}{q^2\, (p^2-m^2_a+i\epsilon)}
\label{se0}
\ee
$\Sigma$ can be computed as a contour integral in the complex plane once the positions of the poles in the $q$ integration variable are determined. From~(\ref{forml}) we get 
\be
P = G^2H^2\,\frac{\mathcal{E}_\gamma}{\sqrt{\en_\gamma^2-m_a^2}} \left(\frac{\sin^2(q_2 L/2)}{q_2^2}+(2\to 1)\right)
\label{rate}
\ee
where the poles in~(\ref{se0}) are located at
\bea
&&q_1=\mathcal{E}_\gamma+p^*\notag\\
&&q_2=\mathcal{E}_\gamma-p^*\notag\\
&&p^*=\sqrt{\mathcal{E}_\gamma^2 -m^2_a}
\label{differen}
\eea
The result~(\ref{rate}) is in perfect agreement with Eq.~(\ref{rate1}). As explicitly shown in the Appendix, this result is obtained {\it only if} $p^*\neq 0$. 

The expression in~(\ref{rate}) includes automatically the case in which the axion recoils backwards with respect to the photon direction -- this is also found in~(\ref{rate1}) provided that  both $p_x>0$ and $p_x<0$ are included when solving the energy $\delta$-function.

We observe now that the wavelength associated to the virtual axion, $\lambda=1/p$, must be of the same size of $L$, for the virtual process to occur within that region\footnote{The uncertainty on the position of the created/destroyed axion is $\Delta x\approx L\Rightarrow \Delta p\gtrsim 1/L$ so that the minimum value of the axion momentum is $p_{\rm min}=0+\Delta p\gtrsim 1/L$.}. Say $\lambda/2< L$  -- the entire wavelength of the virtual axion must be contained in the $H\neq 0$ region. This means that we might estimate $p> 1/2L$. 

Since $p=k-q$ (linear momenta in the $x$ direction), the contribution to $\Sigma$ from forward\footnote{Backward axions have $\en_\gamma-q=-|p|< -1/2L$ so that $q> \en_\gamma+1/2L$.} axions ($p>0$) corresponds to $q<\en_\gamma$ and the above condition on $p$ translates into $q<\en_\gamma-1/2L$ (where $1/2L=10^{-5}$~eV for $L=1$~cm). Similarly for backward ones.  

When $\en_\gamma\neq m_a$, the $q_2$ pole is located within the $q$ region of forward axions, whereas the $q_1$ pole is  in the bouncing-axion region. The poles coincide and are both in the bouncing region when $\en_\gamma=m_a$ (or $p^*=0$). 

In order to estimate a physical  upper bound on $P$, consider $\en_\gamma\neq m_a$ but such that $q_2\lesssim m_a-1/2L$ and $q_1\gtrsim m_a+1/2L$ -- in such a way to maximally approach $m_a$ taking into account the above uncertainty on $p$. This amounts to cut a symmetric segment around $q_1=q_2=m_a$ (we repeat that this specific point does not give contribution to formula~(\ref{rate}) because it has $p^*=0$). 
Therefore, the minimum $q_1-q_2$ distance between the poles in~(\ref{differen}), which still provides the result~(\ref{rate}) is  
\be
\mathrm{min}(q_1-q_2)=1/L
\ee
Confronting with~(\ref{cont}) in the Appendix, we can estimate for forward axions
 \be
P\lesssim G^2H^2 \, \frac{\sin^2(q L/2)}{q^2}\,\frac{m_a}{1/L}
 \ee
where
\be
q\approx m_a-\frac{1}{2L}
\ee
We therefore argue the following formula for forward axions
 \be
 P\simeq G^2H^2 \, \frac{\sin^2(q L/2)}{q^2}\,\frac{\en_\gamma}{1/L+\sqrt{\en_\gamma^2-m_a^2}}
\label{13}
 \ee
with\footnote{We also observe here  that in the differentiation of the integral with respect to $L$, the terms due to the dependency on $L$ of the integration limits do not contribute to the imaginary part of the integral as $\epsilon\to 0$.} $q=q_2=\en_\gamma-\sqrt{\en_\gamma^2-m_a^2}-1/2L$.
As usual, when  $\en_\gamma\gg m_a$ we have $q\approx m_a^2/2\en_\gamma \neq 0$.

This formula will be used to derive the results discussed in the next Section.

\section{The Potential Reach of a sub-THz LSW Experiment}
In this Section, we discuss the potentialities of a Sub-THz-AXion (STAX) search experiment with 
a gyrotron source at 100~kW ($\dot N_\gamma=10^{27}~\gamma/$sec) in a generation Fabry-Perot cavity with quality factor $Q\sim 10^4$. The static  magnetic field is assumed to be $H=15$~T in a region with  $L=50$~cm -- this is the typical length scale over which such intense magnetic
fields may be achieved with available technology. 

From~(\ref{13}) and following equations, we see that, in typical LSW experiment conditions, where $\en_\gamma\gg m_a$, $q$ is a very small number. Therefore the $\sin^2 x/x^2$ function in~(\ref{13}) can be approximated with 1 up to large $L$ values (in the order of several meters).  

\begin{figure}[ht!]
\begin{center}
\includegraphics[width=1.0\columnwidth]{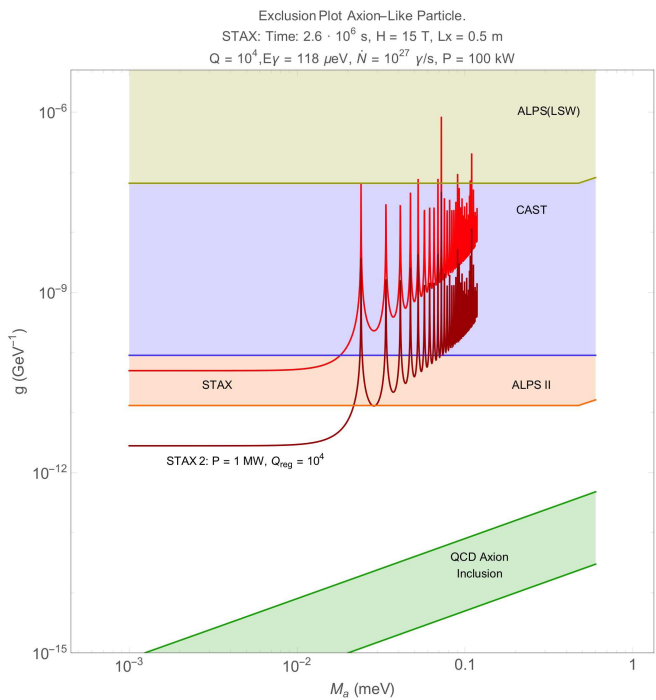}
\end{center}
\caption{\small 90\% CL exclusion limits that STAX may achieve in case of a null result for axions with $m_a\lesssim 0.02$~meV. 
An exposure time of  one month and  zero dark counts are considered.  
`STAX' and `STAX~2' configurations correspond to a 100 kW and 1 MW gyrotron sources respectively. The  former provides $10^{27}$ 
$\gamma$/sec at $30$~GHz. 
}
\label{due}
\end{figure}

For example, in order to observe (or exclude) an axion\footnote{Just below the onset of the oscillation region in Fig.~\ref{due}}  with $m_a\sim 10~\mu$eV from photons with $\en_\gamma\sim 30$~GHz~$=20~\mu$eV, we have to deal with $q\simeq 2.5~\mu$eV whereas for $\en_\gamma\sim 1$~eV,  $q\simeq 5\times 10^{-5}~\mu$eV. On the other hand, a size of $L=50$~cm corresponds to $L\simeq 2.5~\mu$eV$^{-1}$. 

The product $qL/2$ is $qL/2\sim 3$ in the first case and $qL/2\approx 10^{-4}$ in the second case. The function $\sin^2 x/x^2$ can be approximated with $\sim 1$ on very long distances, for visible light, but this is not the case for 30~GHz, where $L$ has to be $L\lesssim 50$~cm. Larger values of $L$ would bring the onset of the oscillation region in Fig.~\ref{due} down to smaller values of the axion mass. This would translate in a smaller sensitivity region in the exclusion plot.

Introducing a Fabry-Perot cavity allows to multiply the passages of the light beam produced by the source in the magnetic field. 
Since  the quality factor $Q$ is proportional to the energy stored in the cavity divided by the energy lost per cycle to the walls, the effective number of passages in the magnetic field gets increased and the LSW probability gains a factor of $Q$ if a cavity is placed on the source side:  $G^4H^4L^4 Q$. 

Therefore, values of $G$ lower by a factor of $Q^{-1/4}$ can be probed in the $G$ ~vs~$m_a$ plot in Fig.~\ref{due}, as commented below.

In Fig.~\ref{due} we show the exclusion plots by ALPS and CAST. ALPS places the strictest limits within purely laboratory experiments while CAST places the strictest limits in astrophysical detection experiments. 

Other projects to be quoted are  OSQAR~\cite{oscar}, IAXO~\cite{iaxo} and ALPS-II~\cite{alps2}. All of them aim at extend the exclusion region in the $M_a$ vs $G$ plot. In particular, ALPS-II can reach a level very close to that of STAX in Fig.~\ref{due} ($G$ down to $2 \times 10^{-11}$~GeV$^{-1}$~\cite{alps2tdr} -- dedicated magnets and optics might help to go towards $10^{-12}$~GeV$^{-1}$~\cite{slides}). The IAXO collaboration aims at decreasing the sensitivity down to $G\sim 10^{-12}$~GeV$^{-1}$ for masses up to about 0.02~eV~\cite{iaxo2}. As for OSQAR, it explores $G$ between  $10^{-7}$ and $10^{-8}$~GeV$^{-1}$.

We superimpose in the same plot the 90\% CL exclusion limits that STAX may achieve in case of a null result. An exposure time of roughly one month and  zero dark counts are considered. 

Since we propose to use sub-THz photons, coherence of photon-to-axion conversion is lost, along with sensitivity, at smaller axion masses compared to other experiments (this is shown by the violent oscillations of the $\sin^2 (q L)/q^2$ function in the conversion probability). 
It is shown in the literature that such oscillations might be reduced introducing a low pressure gas in the generation cavity -- in this case photons acquire an effective mass $m^*\neq 0$ related to the sub-luminal velocity $c/n(k)$  (the in-medium calculations are discussed in~\cite{vbmi}).

The experimental set-up might  further  be improved with the addition of a second Fabry-Perot cavity in the region beyond the wall. As noticed in~\cite{regeneration}, a coherent axion beam may excite electromagnetic modes in a cavity immersed in an external magnetic field: regenerated photons can resonantly enhance at the same cavity frequency, with the effect of producing an increased axion-photon conversion probability by a factor of $Q$ of the cavity (meaning an event probability increased to $G^4H^4L^4Q^2$ or smaller values of $G$ by $Q^{-1/2}$).

In the axion mass range $m_a \lesssim 0.02$~meV, STAX allows an extension of the exclusion region  by a factor of $\sim 10^4$ with respect to ALPS. Adding a second cavity, this gain would be  larger by one more order of magnitude. The ALPS collaboration also foresees a second phase with a regeneration cavity: the ALPS II program. However, STAX II (see STAX II in Fig.~\ref{due}) would still have a better sensitivity, by about a factor of ten.    

The lower  shaded straight band in Fig.~\ref{due} represents the parameter space consistent with the QCD axion from the original Peccei-Quinn theory.

\section{Dark Photons and Chameleons}
The conceptual setup we propose could also be adequate for research on paraphotons and chameleons, as we briefly comment in this section.
In the case of paraphoton searches we might use the same source, same detector but no need of external magnetic field. In the case of chameleons, also the magnetic field in some region $L$ is required, but the experiment is of the after-glow type, as described below.

{\bf  \emph{Paraphotons.}} Paraphotons are hypothetical massive vectors of a hidden U(1) sector, as originally discussed in~\cite{Okun:1982, Holdom:1986}.  For a more recent account, see for example~\cite{Masso:2006gc,Ahlers,Jaeckel:2008}.
The most general Lagrangian with two U(1) gauge groups, describing visible and hidden sector photons and their coupling to visible and hidden matter, can be written as
\bea
	\label{eqn:Lagr01}
	\mathcal L & = & - \frac{1}{4} F^{\mu \nu} F_{\mu \nu} - \frac{1}{4} B^{\mu \nu} B_{\mu \nu} + e J_{\rm em}^\mu A_\mu +\notag\\
	& + & e_h J_{\rm h}^\mu B_\mu - \frac{1}{2} \mu^2 B^\mu B_\mu  
\eea
where $F_{\mu\nu}$ is the field strength tensor for the electromagnetic gauge field, $A^{\mu}$, and $B^{\mu \nu}$ that for the paraphoton, $B^\mu$.
Visible and hidden sector matter currents are $J_{\rm em}^\mu$ and $J_{\rm h}^\mu$ and a mass term for the paraphotons breaks the U(1)$_{\rm h}$ symmetry explicitly. The fields $A$ and $B$ are rotated into $B_1$ and $B_2$ through a small mixing angle $\chi$. $B_{1,2}$ get masses $m_1=\mu\chi$ and $m_2=\mu$ respectively. The value of $\chi$ is expected to  be $\chi \lesssim 10^{-2}$~\cite{Dienes:1997,Goodshell:2009} so that the following approximation $k_1\approx \omega$ is usually made to define the wave vector of $B_1$.

\begin{figure}[htb!]
\includegraphics[width=20pc]{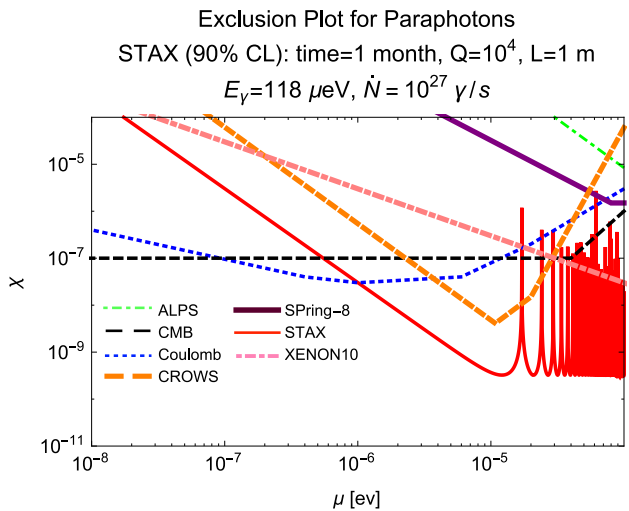} 
\caption{90\% CL exclusion limits that STAX may achieve in case of a null result for paraphotons. An exposure time of one month and zero dark counts are considered.
The STAX parameters are reported on top of the figure. The STAX limits are compared to those provided by ALPS~\cite{ALPS}, CROWS \cite{Betz:2013dza}, SPring-8~\cite{Inada:2013tx} and XENON10 \cite{XENON10} Collaborations, to limits from searches of modifications of Coulomb's law~\cite{bartlett} and constraints on paraphotons from measurements of the CMB~\cite{CMB-COBE-FIRAS}.}
\label{fig:Stax-para-excl}
\end{figure}

\begin{figure}[htb]
\includegraphics[width=16pc]{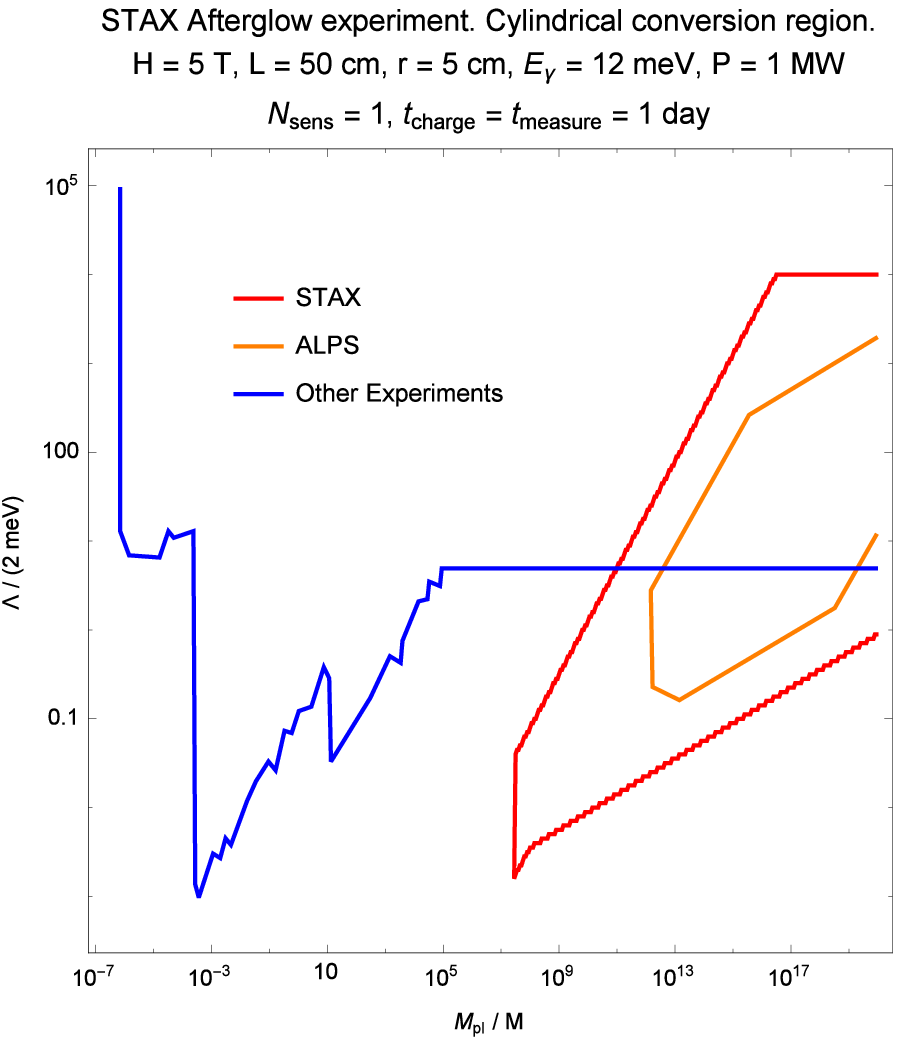}
\caption{Comparison of STAX with other experiments \cite{Mota:2006fz,Ahlers:2007st} at digging exclusion regions for chameleons. The afterglow is considered on a one-day time scale. The STAX parameters and detector sensitivity are reported. $M$ and $\Lambda$ are two theoretical parameters characterizing the chameleons and $M_{pl}$ is the Planck mass. The other parameter in chamaleon theory is $n$, here fixed to $n=1$.}
\label{fig:Stax-cham-excl}
\end{figure}

Since $k_1\neq k_2$, the relative weight of $B_1$ and $B_2$ in the expression for $A(r,t)$ will change with the traversed distance $r$ 
\be
A\propto e^{-i(\omega t -k_1 r)}(B_1+\chi B_2 e^{-i qr})
\ee
with $q=k_1-k_2$. Expressing $B_{1,2}$ in terms of $A,B$ (\mbox{$B_1\propto A-\chi B$} and  \mbox{$B_2\propto \chi A+B$})
we can compute the photon$(A)$-to-paraphoton$(B)$ conversion probability  as
\be
P_{\gamma\to\gamma'}(r) = 4 \chi^2 \mbox{ sin}^2\left(\frac{q\, r}{2}\right)
\ee   
where 
\be
q\simeq\omega-\sqrt{\omega^2-\mu^2}
\ee

The conversion probability can be increased by using a cavity before the wall. If the photon beam is reflected $N$ times, it will make 
$(N+1)/2$ ``attempts'' to cross the wall in a LSW experiment, enhancing the transmission probability by this same
factor. If we define
\be
P_{\rm trans}  =  P_{\gamma\to\gamma'}(L_{1})P_{\gamma'\to\gamma}(L_{2})
\ee
$L_1$ and $L_2$ being the effective traversed lengths of photons and paraphotons,
the expected rate of LSW photons is 
\begin{equation}
	\dot N =  \eta \, \dot N_\gamma \left[ { \frac{N+1}{2}}\right] P_{\rm trans}
\end{equation}
where $\eta$ is the detection efficiency.

In the STAX configuration described above, we take advantage of the very high $\dot N_\gamma$ and quality factor of cavities at sub-THz frequencies.
The single-photon detection with $\eta\approx 1$ will still be assumed. 

As discussed above, low-energy photons from the source might approach the values of the paraphoton masses, so that the approximation $\mu \ll \omega$ is not given for granted.  In this case, $q\approx 0$ and the conversion probability is simply $P=\chi^2 q^2 L^2$, so that it is favorable to have large effective photon/paraphoton paths. Otherwise, as before, we will meet the onset of rapid oscillation at values of $\mu$ approaching $\omega$ -- see Fig.~\ref{fig:Stax-para-excl}. Other collaborations dedicated to the search for photon-paraphotons oscillations in LSW experiments are BMV \cite{BMV}, GammeV \cite{GammeV}, LIPPS \cite{LIPPS}, ALPS \cite{ALPS}, CROWS \cite{Betz:2013dza} and SPring-8 \cite{Inada:2013tx}. In particular, ALPS \cite{ALPS} and SPring-8 \cite{Inada:2013tx} provided the best experimental limits in the range $2\cdot10^{-4} < \mu \lesssim 10^{-3}$ eV and $0.1 < \mu < 0.2$ meV, respectively. CROWS \cite{Betz:2013dza} has recently published updated results, which give the best experimental limit in the region $3\cdot10^{-6} \lesssim \mu \lesssim 3\cdot10^{-5}$ eV. The second stage of ALPS, ALPS-II \cite{alps2}, is planning to extend the previous exclusion region of ALPS for paraphotons up to $\chi \approx 10^{-9}$ for $10^{-4} \lesssim \mu \lesssim 10^{-2}$ eV. See also Ref. \cite{Schwarz:2015lqa}.

{\bf  \emph{Chameleons.}} Chameleons are scalar particles which might have a role at explaining dark energy~\cite{Khoury,Mota:2006fz}. They couple with matter, the electromagnetic field and themselves, through a self-interaction potential. The peculiarity of these particles is the dependence of their mass value on the local mass-energy density.
 
Photons can convert into chameleons when traveling in an external magnetic field, but, differently from the case of axion-like particles, the conversion probability depends on  the square of the energy $\omega^2$ of the photon/chameleon (the dependence on $H$ remains quadratic).  This potentially makes the STAX configuration less suitable for the search of chameleons, for the LSW regenerated photons will have a production probability $\propto \omega^4$ and STAX uses $\sim 20~\mu$eV photons.

Chameleons search experiments are based on a `afterglow' technique \cite{Ahlers:2007st,Gies:2007su} different from axion LSW experiments. Light is injected in a cavity, where a transversal magnetic field is on. Chameleons are supposedly produced inside the cavity, with a certain value of their mass which depends on the density of matter in the hollow part of the cavity. As a result, chameleons might be trapped within the cavity if they miss the necessary amount of energy needed for them to go through the cavity walls: in the denser cavity walls, chameleons are expected to have a larger mass. Low energy STAX photons are more likely producing trapped chameleons. 

Light glows at $\sim 20~\mu$eV should then be detected {\it after} the source has been turned off. The number of chameleons produced in the cavity decreases with an exponential law 
$\exp (-t/ \tau)$ due to chameleon-photon conversion. The time-scale is $\tau=L/P_{\gamma\to \cal{C}}$ where $L$, as usual, is the space extension of the magnetic field and 
\be
 P_{\gamma\to \cal{C}}\approx 4\left( \frac{\omega H}{m^2 M}\right)^2\sin^2\left( \frac{m^2L}{4 \omega}\right)
\ee
is the photon-chameleon conversion probability.  $M$ is a (free) parameter of the chameleon 
model and $m,\omega$ are its mass and energy.

In the low energy regime, where gyrotrons work, the after-glow phenomenon 
might occur on quite long time scales so that the cavity is expected to 
thermalize. Blackbody photons from a hot cavity ($\gtrsim 300$~K) are expected however
to be of minimal impact on the background to 30~GHz afterglow photons.
   
In any case, the photons from source beam must be collimated in such a way to go through 
small size transmitting windows of the cavity to decrease the energy absorption on the body of cavity. If photon-chameleon conversion takes place within the magnetic field in the cavity,
the role of the cavity will only be that of providing a cage to load chameleons in a closed 
region of space.
 
If engineering allows that only $\approx 10^{-3}$ of the the source power heats the body of the cavity, using an hypothetical 1 MW source, we would have a photon background from the hot cavity peaked at $\lambda \approx 80~\mu$m -- three orders of magnitude lower than the typical cm wavelength of after-glow photons at 30 GHz. 

A detailed study on all background sources for the after-glow experiment and on the 
positioning of the light detector will be considered in future work.   

We did not attempt here a full theoretical description of the photon-to-chameleon conversion rate, which is outside the scope of this paper and  can be found in the literature \cite{Ahlers:2007st,Gies:2007su,Chou:2008gr,Rybka:2010ah}. We just report a plot showing the STAX potentiality at searching chameleons when compared to other experiments -- see Fig.~\ref{fig:Stax-cham-excl}. 

The other exclusion limits reported in Fig.~\ref{fig:Stax-cham-excl} are those of the ALPS Collaboration, Alpenglow experiment \cite{Ahlers:2007st} (orange line), and constraints arising from searches for violations of the weak equivalence principle, from searches for deviations from the $1/r^2$ law of the gravitational force, and from bounds on the strength of any fifth force from measurements of the Casimir force (blue line) \cite{Mota:2006fz}.

\section{Single-photon detection of sub-THz radiation}
We argue that a suitably designed superconducting antenna connected to a nano-sized hot-electron calorimeter read out by an ultra low noise superconducting quantum interference device (SQUID)  amplifier can provide the desired sensitivity   to detect   even a  single photon from an axion-to-photon conversion.
 
 The calorimeter would consist  of an absorbing metallic nano-wire made of a  superconducting material operating near to the superconducting-to-normal state phase transition.  The photon absorption will cause  the temperature of the metal electrons to rise, thereby inducing an enhancement of the calorimeter resistance. This kind of detector is usually known as transition edge sensor (TES). Calorimeters based on TES have been successfully exploited as single-photon quantum detectors for X-ray spectroscopy~\cite{bolo5} as well as for secure quantum communication applications using near-IR photons \cite{bolo6}. Nano-calorimeters are able to extend the ability of conventional calorimeters to detect both very tiny powers as well as small amounts of energy which potentially correspond to a single sub-THz or microwave photon event.

The TES calorimeter  sensitivity is basically limited by  the magnitude of  the thermal energy fluctuations, due to the energy exchange between the sensor and the phonon bath. The RMS energy resolution $\sigma_E$  can be written as $\sigma_E \sim  0.3 \sqrt{k_B T_c^2 C_e}$  \cite{bolo7,bolo8}, where  $k_B$ is the Boltzmann  constant, $T_c$   the critical temperature of the TES superconductor and $C_e$ the  heat capacitance. In addition, $C_e$  is given by  $\gamma V T_e$, where  $\gamma$ is the Sommerfeld coefficient\footnote{ The Sommerfeld coefficient $\gamma$ is the ratio of the electronic specifc heat to temperature T. It is determined to be $\gamma = n(\pi k_B)^2/(2\epsilon_F)$, where $n$ is the electron gas charge density, $\epsilon_F$ is the Fermi energy, and $k_B$ the Boltzmann constant.
},   $V$ is the sensor volume and $T_e$ the electron temperature. The above expression suggests that  low  $T_e$ and a reduced sensor volume are required to detect low-energy photons. 

By using, for instance, $\alpha$-tungsten  as TES material  with a critical temperature of  $\sim  15$~mK,  $\gamma$  $\sim 136 $~J/K/m$^3$, and  $ V= 300\times 40\times 20 $~nm$^3$,  we expect a frequency resolution   $\sigma_E $ $\sim$ 560 MHz and therefore  we predict a significant resolving power $h\nu/\sigma_E\sim$ 50-180  for 30-100 GHz photons~\cite{bolo9}. In this picture, the dark count noise is well under control and the quantum efficiency of the device maximal.

The photon energy can  be fed into the TES through a highly-efficient log-periodic spiral antenna, with a suitable size to match the photon wavelength,   to be devised  with superconductor  materials which are in direct contact with the TES. The antenna will be designed to obtain the best possible impedance matching with the vacuum.

With a suitable choice of the TES material, a tiny but detectable current variation will be obtained. A  temperature variation of the TES  of several tens of mK  might be achieved together with an electron-phonon coupling of the nano-wire of the order of few aW/K~\cite{bolo9}. An  expected signal of $\sim 10^{-20}$ W would translate into a current variation of some tens pA with a suitable bias voltage applied to the TES. Therefore, ultra low-noise DC SQUID readout amplifiers, inductively coupled to the TES circuit, are  required. The best current sensitivity with SQUIDs reported to date (4 fA/$\sqrt{\rm Hz}$)~\cite{SQUID},  together with a reduced bandwidth of few kHz, will result into a total current noise induced by the SQUID which is two orders of magnitude smaller  than the expected current amplitude of the TES response.
 
  This quantum sensor will be operated in a dilution refrigerator at a temperature of 10~mK  properly coupled to the conversion magnetic field. Blackbody radiation in the refrigerator should induce a negligible background in the range of interest (30 GHz to 100 GHz): the  peak of the blackbody spectrum is at a value of  $\approx 0.6$~GHz, much lower than the working frequency. In an energy range of $\pm~10$~GHz around the central value of 30~GHz, the irradiance is calculated to be $10^{-52}$~W/m$^2$ with  $10^{-30}$~m$^{-2}$s$^{-1}$ emitted photons.   Other sources  of energy deposit  in the absorber --   as cosmic ray radiation or radioactivity  -- should produce a very large signal: cosmic muons release about 10~eV in 10~nm of material.  This would drastically increase the temperature of the detector and a  dead-time  would be needed to get back to the working temperature of 10~mK. We estimate  this time scale to be much lower than the average time interval between two cosmic muon events hitting the detector.

\section{Conclusions}
We have presented the conceptual proposal of  a different kind of Light-Shining-through-Wall experiment based on low energy, very intense photon sources (30~GHz) and very sensible ultra-cold single photon detectors. 

We have discussed the photon-to-axion conversion probability in this photon frequency regime and showed the potentialities of the configuration we propose at digging the exclusion region for axion-like particles in a very competitive way with respect to other LSW experiments. We have briefly commented on the kind of detectors that could be devised for our purposes. 

A brief analysis on the potentialities for the search of  paraphotons and chameleons is also included. 

The concept experiment  proposed is found to be especially effective at axion-like particles searches whereas it has similar potentialities of other running or concept experiments in the case of paraphotons. As for chameleons, the exclusion region is extended  in a sector of the parameter space which will not be covered by future space-based tests~\cite{Ahlers:2007st}.
 
\subsection*{Acknowledgements}
We whish to thank Andreas Ringwald for comments on the manuscript and especially for pointing us about some studies we were not aware of on the use of gyrotrons in LSW experiments.
We thank Marcella Diemoz for encouraging this research activity at INFN Rome. 

\section*{Appendix}
The Feynman rules to compute $\Sigma(k)$  in Fig.~\ref{uno} are the following. At each vertex attach (see~(\ref{frule2}))
\bea
&&(2\pi)^3\delta(k_y-p_y)\delta(k_z-p_z)\delta(\en_\gamma-\en_a)(2\pi)\delta(k_x-p_x-q_x)\notag\\
&&(2\pi)^3\delta(p_y-k_y^\prime)\delta(p_z-k_z^\prime)\delta(\en_a-\en_\gamma^\prime) (2\pi)\delta(p_x+q_x-k_x^\prime)\notag
\eea
where the last two delta-fuctions substitute the action of the external magnetic field. Integrate over both internal lines with the prescription  
\be
\frac{d^4p}{(2\pi)^4} \frac{i\, dq_x}{2\pi}
\ee
considering that there are no transverse components.
 The first integration allows to factor out the term
\be
(2\pi)^4\delta^4(k-k^\prime)
\ee
thus, including the remaining terms 
in~(\ref{frule2})  we get
\be
\Sigma(k_x)=-4\,G^2 H^2\,\mathcal{E}_\gamma^2\, \int\,\frac{dq_x}{2\pi} \frac{\sin^2(q_x L_x/2)}{q_x^2\, (p^2-m^2_a+i\epsilon)}
\label{se}
\ee
where we used the  propagator for an axion of mass $m_a$ as given by
$\Delta(p)= i/(p^2-m^2_a+i\epsilon)$.

A factor of $1/2!$ from the second perturbative order has been included, together with a factor of 2 coming from the fact that the two vertices can be interchanged. The overall $(-1)$ factor in~(\ref{se})  reflects the fact that the photon is incoming at one vertex and outgoing at the other. From now on let us simplify the notation by setting $L_x=L$, $p_x=p$, $k_x=k$, $q_x=q$. 

The $ (p^2-m^2_a+i\epsilon)$ term in the denominator in~(\ref{se}) can be written as
\bea
&&(p^2-m^2_a+i\epsilon)= (\mathcal{E}_\gamma^2-(k-q)^2-m^2_a+i\epsilon)=\notag\\
&&=-(q-(q_1+i\epsilon))(q-(q_2-i\epsilon))
\eea 
where in the last term we redefined 
\be
i\epsilon\rightarrow i\epsilon \,p^* - \epsilon^2
\ee
with $\epsilon\to 0$. To do this it is necessarily required that $p^*\neq 0$. 
We also assumed $k=\mathcal{E}_\gamma>m_a $ and defined the two positive roots
\bea
&&q_1=\mathcal{E}_\gamma+p^*\notag\\
&&q_2=\mathcal{E}_\gamma-p^*
\label{differ}
\eea
in such a way that $q_1-q_2>0$, and
\be
p^*=\sqrt{\mathcal{E}_\gamma^2 -m^2_a}
\label{cinem}
\ee
The $q_1-q_2>0$ condition is essential to define the prescription to lift the $q_{1,2}$ poles in the complex plane and is lost when $p^*= 0$. 

Next we observe that taking a derivative of  the integral in~(\ref{se}) with respect to $L$ we obtain,
  leaving aside constant factors for the moment
\bea
&&\frac{1}{2}\int\,dq \frac{\sin(q L)}{q\, (p^2-m^2_a+i\epsilon)}=\notag\\
&&=-\frac{1}{2}\int\,dq \frac{\sin(q L)}{q\, (q-(q_1+i\epsilon))(q-(q_2-i\epsilon))}
\label{se3}
\eea
The last integral can be evaluated as a contour integral in the complex plane and we find that
\bea
&&\int_{-\infty}^{\infty}dx\;\frac{\sin(\alpha x)}{x(x-(a+i\epsilon))(x-(b-i\epsilon))} =\notag\\
&&= \frac{\pi}{ab}+\frac{\pi}{a-b}\left(\frac{e^{i\alpha a}}{a}-\frac{e^{-i\alpha b}}{b}\right)
\label{cont}
\eea
for $a,b$ real parameters.

Comparing with~(\ref{se3}) we have $\alpha=L$, $a=q_1$, $b=q_2$.  Therefore the result of the integral in~(\ref{se3}), upon integration over $L$ is
\be
-\frac{\pi L}{2q_1q_2}+\frac{i\pi}{4p^*}\left(\frac{e^{iLq_1}-1}{q_1^2}+\frac{e^{iLq_2}-1}{q_2^2}\right)
\label{partres}
\ee
where the integration constant of the indefinite integration has been fixed requiring the physical condition $\Sigma\to 0$ as $L\to 0$. 

The imaginary part of~(\ref{partres}) is therefore
\be
-\frac{\pi}{2p^*}\left(\frac{\sin^2(q_1L/2)}{q_1^2}+\frac{\sin^2(q_2L/2)}{q_2^2}\right)
\label{risul}
\ee

From the unitarity of the S-matrix (cutting-equations) we conclude that the width of the unstable photon in the external magnetic field (conversion rate of a real photon into an axion) is  twice the imaginary part of $\Sigma$. 

From Eq.~(\ref{forml}), (\ref{se}) and~(\ref{risul}), and including the appropriate wave function normalization $1/(\sqrt{2\en_\gamma})^2$,  
we have therefore, as long as $p^*\neq0$
\be
P = G^2H^2\,\frac{\mathcal{E}_\gamma}{\sqrt{\en_\gamma^2-m_a^2}} \left(\frac{\sin^2(q_2 L/2)}{q_2^2}+(2\to 1)\right)
\ee
in  agreement with Eq.~(\ref{rate1}). Notice that if we let $\alpha\to\infty$ in the lhs of Eq.~(\ref{cont}), the imaginary part will be zero in the $\epsilon\to 0$ limit. This agrees with the discussion preceding Eq.~(\ref{rate1}).



\begin{thebibliography}{99}

\bibitem{gen} 
See for example: 
  J.~Jaeckel and A.~Ringwald, Ann. Rev. Nucl. Part. Sci. {\bf 60}, 405 (2010); 
  A.~Ringwald, Phys. Dark. Univ. {\bf 1}, 116 (2012); 
  J.~Hewett {\it et al.}, arXiv:1205.2671; 
  S.~Andreas, C.~Niebuhr and A.~Ringwald, Phys. Rev. D {\bf 86}, 095019 (2012); 
  S.~Andreas, M.~D.~Goodsell and A.~Ringwald, Phys.\ Rev.\ D {\bf 87}, no. 2, 025007 (2013).

As for older seminal contributions, see:
  P.~Sikivie, Phys. Rev. Lett. {\bf 51}, 1415 (1983); Phys. Rev. D {\bf 32}, 2988 (1985); 
  G.~Raffelt and L.~Stodolsky, Phys. Rev. D {\bf 37}, 1237 (1988).

\bibitem{Arik:2015cjv}
  M.~Arik {\it et al.} [CAST Collaboration],
  Phys.\ Rev.\ D {\bf 92} (2015) 2,  021101
  [arXiv:1503.00610 [hep-ex]].

\bibitem{crq} R.N.~Clarke and C.B. Rosenberg, {\it Fabry-Perot and
open resonators at microwave and millimetre wave frequencies, 2-300 GHz}, 
J. Phys. E: Sci. Instrum., Vol. 15, (1982), see section 1.2.  

\bibitem{japgyr}
LSW from gyrotrons is pursued by a Japanese group at the University of Tokyo, although in a different energy range from the one discussed in this paper. In particular, see the slides {\tt http://tabletop.icepp.s.u-tokyo.ac.jp/Tabletop\newline$\_$experiments/WISP$\_$search$\_$with$\_$sub-THz$\_$photon$\_$files\newline/suehara-irmmw-2012.pdf} and the conference proceedings by T.~Suehara, K.~Owada,  A.~Miyazaki, T.~Yamazaki, S. Asai, T.~Kobayashi, ``Hidden particle search using Sub-THz gyrotron", in IRMMW-THz 2012, 37th International Conference on Infrared, Millimeter, and Terahertz Waves, University of Wollongong, Australia, 23-28 Sept. 2012, pp. 1-2, {\tt http://ieeexplore.ieee.org/xpl/articleDetails.jsp\newline ?reload=true\&arnumber=6380414}.

\bibitem{weinbergokun} 
  S. Weinberg, {\it The Quantum Theory of Fields}, vol. I, Cambridge (2005), see chapter on $S$-Matrix; 
  L. Okun, {\it Quarks and Leptons}, North Holland (1985),  see p.~326.


\bibitem{vanbibbr} 
  K.~Van Bibber, N.~R.~Dagdeviren, S.~E.~Koonin, A.~Kerman and H.~N.~Nelson,
  Phys.\ Rev.\ Lett.\  {\bf 59}, 759 (1987).

\bibitem{oscar}
  R.~Ballou {\it et al.} [OSQAR Collaboration],
  arXiv:1506.08082 [hep-ex].

\bibitem{iaxo} 
  E.~Armengaud {\it et al.} [IAXO Collaboration],
  JINST {\bf 9}, T05002 (2014).
  
\bibitem{alps2}
  R.~B\"ahre {\it et al.} [ALPS-II Collaboration],
  JINST {\bf 8}, T09001 (2013);
  B.~D\"obrich [ALPS-II Collaboration],
  arXiv:1309.3965 [physics.ins-det].
  
\bibitem{alps2tdr}
  See the ALPS-II TDR at \url{http://arxiv.org/abs/arXiv:1302.5647}.
 
\bibitem{slides}  
 {\tt http://desy.de/~ringwald/axions/talks/ultralight\newline$\_$frontier$\_$mainz.pdf}.
 
 \bibitem{iaxo2} 
  T.~Dafni {\it et al.} [IAXO and CAST Collaborations],
  PoS TIPP {\bf 2014}, 130 (2014). 
  
\bibitem{vbmi} 
  K.~van Bibber, P.~M.~McIntyre, D.~E.~Morris and G.~G.~Raffelt,
  Phys.\ Rev.\ D {\bf 39}, 2089 (1989).
  
\bibitem{regeneration} 
  P.~Sikivie, D.B.~Tanner, K.~van Bibber, 
  Phys. Rev. Lett. {\bf 98}, 172002 (2007); 
  G.~Mueller, P.~Sikivie, D.~B.~Tanner and K.~van Bibber,
  Phys. Rev. D {\bf 80}, 072004 (2009).

\bibitem{Okun:1982}
  L. B. Okun,
  Zh. Eksp. Teor. Fiz. {\bf 83}, 892 (1982) 
  [Sov. Phys. JETP {\bf 56}, 502 (1982)].
  
  \bibitem{Holdom:1986}
  B. Holdom,
  Phys. Lett. B {\bf 166}, 196 (1986).
  
  \bibitem{Masso:2006gc}
  E.~Mass\'o and J.~Redondo,
  Phys.\ Rev.\ Lett.\  {\bf 97} (2006) 151802.

\bibitem{Ahlers}
  M. Ahlers, H. Gies, J. Jaeckel, J. Redondo and A. Ringwald,
  Phys.\ Rev.\ D {\bf76}, 115005 (2007).
  
\bibitem{Jaeckel:2008}
  J. Jaeckel and A. Ringwald,
  Phys. Lett. B {\bf 659}, 509 (2008).     
  
 \bibitem{Dienes:1997}
  K. R. Dienes, C. F. Kolda and J. March-Russell,
  Nucl. Phys. B {\bf 492}, 104 (1997).
  
\bibitem{Goodshell:2009}
  M. Goodshell, J. Jaeckel, J. Redondo and A. Ringwald,
  JHEP {\bf 0911}, 027 (2009).        
  
  \bibitem{ALPS}
  K. Ehret {\it et al.} [ALPS Collaboration],
  Phys. Lett. B {\bf 689}, 149 (2010).
  
\bibitem{Betz:2013dza} 
  M.~Betz, F.~Caspers, M.~Gasior, M.~Thumm and S.~W.~Rieger [CROWS Collaboration],
  Phys.\ Rev.\ D {\bf 88}, no. 7, 075014 (2013).  
  
\bibitem{Inada:2013tx} 
  T.~Inada, T.~Namba, S.~Asai, T.~Kobayashi, Y.~Tanaka, K.~Tamasaku, K.~Sawada and T.~Ishikawa,
  Phys.\ Lett.\ B {\bf 722}, 301 (2013).
  
\bibitem{XENON10}
  J. Angle {\it et al.} [XENON10 Collaboration],
  Phys. Rev. Lett. {\bf 107}, 051301 (2011);	
  H. An, M. Pospelov and J. Pradler, 
  Phys. Rev. Lett. {\bf 111}, 041302 (2013).  
  
  \bibitem{bartlett}
  D. F. Bartlett and S. Loegl,
  Phys. Rev. Lett. {\bf 61}, 2285 (1988).              
  
\bibitem{CMB-COBE-FIRAS}
  D. J. Fixsen {\it et al.},
  Astrophys. J. {\bf 473}, 576 (1996).	   
  
\bibitem{BMV}
  M. Fouch\'e {\it et al.} [BMV Collaboration], 
  Phys. Rev. D {\bf 78}, 032013 (2008).  
  
\bibitem{GammeV}
  A. Chou {\it et al.} [GammeV Collaboration], 
  Phys. Rev. Lett. {\bf 100}, 080402 (2008).  
  
\bibitem{LIPPS}
  A. Afanasev {\it et al.} [LIPPS Collaboration], 
  Phys. Rev. Lett. {\bf 101}, 120401, (2008);
  Phys. Lett. B {\bf 679}, 317 (2009).    
  
\bibitem{Schwarz:2015lqa} 
  M.~Schwarz, E.~A.~Knabbe, A.~Lindner, J.~Redondo, A.~Ringwald, M.~Schneide, J.~Susol and G.~Wiedemann,
  JCAP {\bf 1508}, no. 08, 011 (2015).  
  
\bibitem{Khoury}
  J. Khoury and A. Weltman, 
  Phys. Rev. D {\bf69}, 044026 (2004);  
  Ph. Brax, C. van de Bruck, A.-C. Davis, J. Khoury and A. Weltman, 
  Phys. Rev. D {\bf70}, 123518 (2004). 
  
\bibitem{Mota:2006fz} 
  D.~F.~Mota and D.~J.~Shaw,
  Phys.\ Rev.\ D {\bf 75}, 063501 (2007).

\bibitem{Ahlers:2007st} 
  M.~Ahlers, A.~Lindner, A.~Ringwald, L.~Schrempp and C.~Weniger,
  Phys.\ Rev.\ D {\bf 77}, 015018 (2008).

\bibitem{Gies:2007su} 
  H.~Gies, D.~F.~Mota and D.~J.~Shaw,
  Phys.\ Rev.\ D {\bf 77}, 025016 (2008).    
  
\bibitem{Chou:2008gr}
  A. S. Chou {\it et al.} [GammeV Collaboration],
  Phys. Rev. Lett. {\bf 102}, 030402 (2009);
  A.~Upadhye, J.~H.~Steffen and A.~Weltman,
  Phys.\ Rev.\ D {\bf 81}, 015013 (2010). 
  
\bibitem{Rybka:2010ah} 
  G.~Rybka {\it et al.} [ADMX Collaboration],
  Phys.\ Rev.\ Lett.\  {\bf 105}, 051801 (2010).  
  
\bibitem{bolo5} 
   K. D. Irwin and G. C. Hilton, 
   Cryogenic Particle Detection, vol.  99, ed. Berlin: Springer-Verlag Berlin, p. 63 (2005).
   
 \bibitem{bolo6}  A. J. Miller, S. W. Nam, J. M. Martinis, and A. V. Sergienko, 
   Appl. Phys. Lett. {\bf 83}, 791 (2003).

 \bibitem{bolo7}  
   K. D. Irwin, ``Phonon-mediated particle detection using superconducting tungsten transition-edge sensors", 
   PhD thesis, Department of Physics, Stanford University, 1995. 
   
\bibitem{bolo8} 
  B. S. Karasik and A.V. Sergeev, 
  IEEE Trans. Appl. Supercond. {\bf 15}, 618 (2005).
  
\bibitem{bolo9}
  F. Giazotto, T. T. Heikkila, A. Luukanen, A. M. Savin and J. P. Pekola,
  Rev. Mod. Phys. {\bf 78}, 217 (2006).  

\bibitem{SQUID}
  F.~Gay, F.~Piquemal and G.~Geneves 
  Rev. Sci. Instrum. {\bf 71}, 4592 (2000).  

\end{thebibliography}
\end{document}